# Practical Link Adaptation Algorithm with Power Density Offsets for 5G Uplink Channels

Shu Sun, Sungho Moon, and Jong-Kae Fwu

*Abstract*—This letter proposes a pragmatic link adaptation algorithm considering power density offsets (PDOs) for next-generation uplink wireless channels. The proposed algorithm consists of PDO calculation between a physical uplink shared channel and its associated sounding reference signal, key channel state metric generation, and modulation and coding scheme (MCS) adaptation with respect to the PDO. Scaling is applied to estimated channel matrices based on multiple reference PDO points to generate corresponding reference mutual information (MI) values, followed by interpolation or extrapolation to obtain the adapted MI and ultimately MCS. The proposed algorithm has low complexity in terms of hardware implementation, while yielding satisfactory block error rates and throughput for a wide range of PDOs as shown by simulation results.

## I. Introduction

Fifth-generation (5G) wireless communication techniques are expected to provide orders of magnitude higher data rates and reliability compared to previous generations. In order to establish and retain a reliable radio link with satisfactory throughput and block error rate (BLER) between a next-generation NodeB (gNB) and its user equipment (UE), the transmitter, i.e., gNB in the downlink and UE in the uplink, should transmit data with a proper modulation and coding scheme (MCS) according to channel conditions. The MCS is usually computed via channel state information (CSI) feedback obtained using reference signals. The 5G new radio (NR) specifications do not strictly specify the approach to MCS selection [1]-[5], but usually a technique is employed where the MCS, which achieves highest data rate (i.e., maximum transport block size) while maintaining a target BLER, is selected. If the CSI is not accurate due to various reasons such as aging, channel fading, and transmit power change, the MCS needs to be adapted accordingly.

For the physical uplink shared channel (PUSCH) in 5G NR [1], the MCS that a UE should transmit with is obtained through the CSI acquired using a sounding reference signal (SRS). There exists rich literature on outer-loop link adaptation that utilizes acknowledgment/negative-acknowledgment results to adjust the MCS [6]-[10], or for inner-loop link adaptation based on estimated packet error rate [11] or considering CSI aging [12]. To the best knowledge of the authors, however, no previous solutions are available in the literature to solve the problem considering transmit power density offsets (PDOs) between PUSCH and SRS, where the PDO refers to the power per resource element (RE).

In this letter, we propose a link adaptation algorithm for PUSCH when there is a transmit PDO between the PUSCH and its associated SRS. The estimated channel matrix from SRS is scaled using multiple reference PDO values to obtain the corresponding mutual information (MI) per resource block group (RBG), resulting in an MI curve with multiple points. Given a PDO between PUSCH and SRS at any moment, the corresponding MI per RBG can be obtained via interpolation or extrapolation of the MI vs. PDO curve, and the corresponding MI per RBG will be used to schedule uplink transmission by selecting a more accurate MCS via an MI-to-MCS mapping (MI2MCS) table based on a target BLER. A method is also proposed to select a rank indicator (RI) and a transmit precoding matrix index (TPMI) required for uplink scheduling.

## II. Proposed Link Adaptation Algorithm

The PDO between PUSCH and its associated SRS can be dynamically varying due to timing relationship between PUSCH and SRS, as well as power limitations of the UE, which will be detailed in the following subsection. This dynamic variation in PDO necessitates a link adaptation algorithm to adapt the MCS for PUSCH transmission in a timely manner.

### A. Power Density Offset Calculation

The transmit power of PUSCH and SRS are specified in the 5G NR technical specification (TS) 38.213 by the 3rd Generation Partnership Project (3GPP) [3]. The PUSCH transmit power $P_{\text{PUSCH}}$ is given by Eq. (1) (with an abridgement of subscripts and indices for more concise formulation hence easier understanding) [3], where $P_{\text{CMAX}}$ represents the UE

The authors are with the Next Generation and Standards (NGS) group of Intel Corporation, Santa Clara, CA 95054 USA (e-mail: shu.sun@intel.com; sungho.moon@intel.com; jong-kae.fwu@intel.com).

$$P_{PUSCH} = min \begin{Bmatrix} P_{CMAX} \\ P_{O\_PUSCH} + 10log_{10}(2^\mu \times M_{RB}^{PUSCH}) + \alpha_{PUSCH} \times PL + \Delta_{TF} + f \end{Bmatrix} [dBm] \quad (1)$$

$$P_{SRS} = min \begin{Bmatrix} P_{CMAX} \\ P_{O\_SRS} + 10log_{10}(2^\mu \times M_{RB}^{SRS}) + \alpha_{SRS} \times PL + h \end{Bmatrix} [dBm] \quad (2)$$

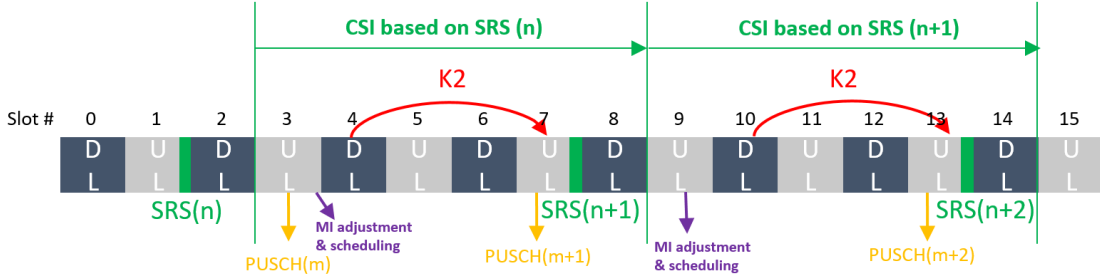

Fig. 1. Example of timing relationship of SRS and PUSCH. DL and UL represent downlink and uplink, respectively. K2 denotes the slot offset from downlink transmission of PUSCH scheduling information to the associated PUSCH transmission.

configured maximum output power, $P_{O\_PUSCH}$ is the noise adjustment parameter for PUSCH, $\mu$ represents the subcarrier spacing configuration [1], $M_{RB}^{PUSCH}$ denotes the bandwidth of the PUSCH resource assignment expressed in number of resource blocks (RBs), $\alpha_{PUSCH}$ is the power control parameter for PUSCH, PL stands for path loss, $\Delta_{TF}$ is an adjustment factor to account for different modulation and coding, and $f$ denotes the PUSCH power control adjustment state [3]. The SRS transmit power $P_{SRS}$ is shown by Eq. (2) [3] (with an abridgement of subscripts and indices), in which $P_{O\_SRS}$ is the noise adjustment parameter for SRS, $M_{RB}^{SRS}$ denotes the SRS bandwidth expressed in number of RBs, $\alpha_{SRS}$ is the power control parameter for SRS, and $h$ is a factor related to power control adjustment state [3]. Based on Eqs. (1) and (2), the power densities of PUSCH and SRS, $S_{PUSCH}$ and $S_{SRS}$, can be expressed as Eqs. (3) and (4), respectively:

$$S_{PUSCH} = P_{PUSCH} - 10log_{10}(M_{RB}^{PUSCH} \times M_{RE}^{RB,PUSCH}) \quad (3)$$

$$S_{SRS} = P_{SRS} - 10log_{10}(M_{RB}^{SRS} \times M_{RE}^{RB,SRS}) \quad (4)$$

where $M_{RE}^{RB,PUSCH}$ and $M_{RE}^{RB,SRS}$ denote the number of occupied REs per RB for PUSCH and SRS, respectively. Eqs. (1) and (2) indicate that transmit powers of PUSCH and SRS depend on the number of RBs. For SRS, the entire bandwidth (e.g., 272 RBs for sub-6 GHz carriers) is usually used to obtain adequate CSI over frequency resources, thus $P_{CMAX}$ is likely to be met in many occasions, leading to lower $S_{SRS}$. Moreover, $M_{RE}^{RB,PUSCH}$ and $M_{RE}^{RB,SRS}$ may also differ from each other, further contributing to the discrepancy of $S_{PUSCH}$ and $S_{SRS}$.

The PDO is calculated as the difference between $S_{PUSCH}$ and $S_{SRS}$ corresponding to the latest PUSCH available and the latest SRS available when performing PUSCH scheduling at a gNB, where the expected transmit power density is based upon power control command from higher layers. Fig. 1 illustrates an example of timing relations of SRS and PUSCH, where DL and UL represent downlink and uplink, respectively, and K2 (which equals three in Fig. 1) denotes the slot offset from downlink transmission of PUSCH scheduling information to the associated PUSCH transmission. For instance, the scheduling decision for PUSCH(m+1) in Slot 7 is made before the end of Slot 3, by which moment the latest available PUSCH and SRS powers correspond to PUSCH(m) and SRS(n). Therefore, the PDO is calculated based on the transmit power of PUSCH(m) and SRS(n) and their frequency allocations.

### B. RI, TPMI, and MI Generation from SRS

RI, TPMI, and MI per RBG are indispensable to scheduling PUSCH transmission by signaling on a physical downlink control channel (PDCCH) [2], hence the acquirement of these parameters is critical. At the gNB receiver side, after obtaining the estimated channel matrix $\boldsymbol{H}_{est}$ per every RBG based on wideband (WB) SRS sent by a UE, Algorithm 1 is conducted to generate RI, TPMI, and MI. An RBG refers to a set of consecutive virtual RBs defined by higher layers, and an RB is defined as 12 consecutive subcarriers in the frequency domain [1]. WB refers to the radio band including all the RBGs in the system bandwidth. The RI ranges from 1 to the minimum of the numbers of transmit and receive antenna ports. In Algorithm 1, $\boldsymbol{W}$ denotes the noise whitening matrix, and $\boldsymbol{P}(RI, TPMI)$ represents the candidate precoding matrix as a function of RI and TPMI [1]. $N_{RI}, N_{TPMI}, N_{RBG}, N_{modulation}$ denote the total number of RI, TPMI, RBG, and modulation schemes, respectively. The selected optimum WB RI and WB TPMI are $r_{OPT}$ and $p_{OPT}$, respectively, which maximizes the WB MI, while $MI(r_{OPT}, p_{OPT}, b)$ is the optimum MI for RBG b. Essentially, an MI dictionary with a hierarchical architecture is generated as a function of RI, TPMI, RBG, and modulation scheme (Step 6 in Algorithm 1), then a maximum MI is selected over all modulation schemes for each RBG, which are stored and averaged over all RBGs to yield the WB MI as a function of RI and TPMI. Afterwards, the optimal combination of RI and TPMI is selected that renders the maximum WB MI, and the corresponding RI and TMPI are output as the final WB RI and WB TPMI. Eventually, for each RBG, the MI associated with the WB RI and WB TPMI is selected as the optimal MI.

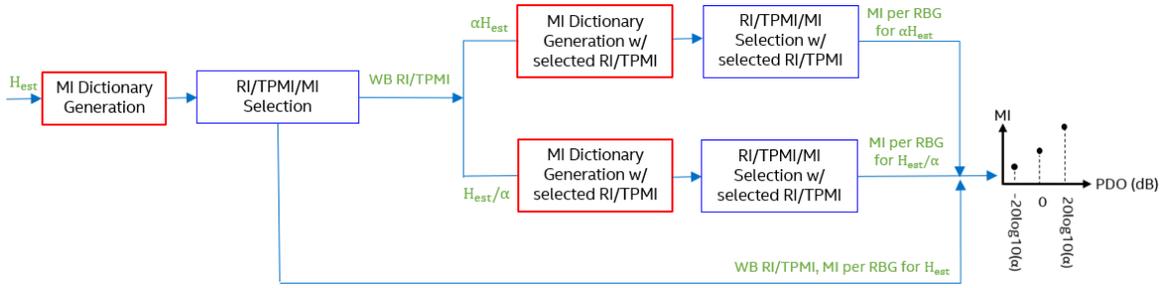
Fig. 2. An example of the block diagram for generating WB RI, WB TPMI, and MI per RBG as a function of PDO.

## C. Generation of MI as A Function of PDO

To produce MIs corresponding to various PDOs, scaling is applied to $\mathbf{H}_{est}$, resulting in two scaled estimated channel matrices $\alpha \mathbf{H}_{est}$ and $\mathbf{H}_{est}/\alpha$, where two virtual PDO values between PUSCH and SRS, $20\log_{10}(\alpha)$ dB and $-20\log_{10}(\alpha)$ dB, are assumed and utilized as reference points. Using $\mathbf{H}_{est}$, $\alpha \mathbf{H}_{est}$, and $\mathbf{H}_{est}/\alpha$, as well as Algorithm 1, we can obtain three MI values corresponding to reference PDO values of 0 dB, $20\log_{10}(\alpha)$ dB, and $-20\log_{10}(\alpha)$ dB, respectively. Note that the WB RI and WB TPMI are derived from the unscaled $\mathbf{H}_{est}$. For the scaled estimated channel matrices $\alpha \mathbf{H}_{est}$ and $\mathbf{H}_{est}/\alpha$, Steps 9-11 in Algorithm 1, is performed with the RI and TPMI generated using the unscaled $\mathbf{H}_{est}$, instead of with all candidate RIs and TPMIs, which significantly reduces hardware complexity and computation time due to parallel processing. Fig. 2 illustrates an example of the block diagram for generating WB RI, WB TPMI, and MI per RBG as a function of PDO. It is noteworthy that variants of the architecture in Fig. 2 can be implemented in practice.

## D. Adaptive MCS Generation

The MI associated with an actual (as opposed to reference) PDO is calculated via interpolation or extrapolation of the MI vs. PDO curve depicted by the rightmost example plot in Fig. 2. Finally, the MCS is obtained through an existing MI2MCS table calibrated to yield around 10% BLER. The above process is done for each RBG and each UE to compute the MCS per RBG per UE. According to our analysis, the PDO between PUSCH and SRS can range from -20 dB to +20 dB. Therefore, the reference PDO value $20\log_{10}(\alpha)$ is set to a middle point of 10 dB in the simulations shown later, i.e., $\alpha = \sqrt{10}$. Nevertheless. simulation results are not sensitive to the selection of $\alpha$ based on our observations. The overall processes presented in Sections II-C and II-D are summarized in Algorithm 2.

## III. SIMULATION RESULTS

Simulations are performed to demonstrate the viability and effectiveness of our link adaptation algorithm. Simulation settings are given in Table I. The comparison of BLERs and

---

**Algorithm 1: RI, TPMI, and MI Generation Algorithm**

**Input:** $H_{est}$ per RBG, $W$ per RBG, and $P$(RI, TPMI)

**Output:** WB RI, WB TPMI, MI per RBG

1. **for** $r = 1 : N_{RI}$
2.     **for** $p = 1 : N_{TPMI}$
3.         **for** $b = 1 : N_{RBG}$ **do**
4.             Compute $\widetilde{H}(r,p,b) = W(b) H_{est}(b) P(r,p)$,
   Compute SINR$(r,p,b)$ based on $\widetilde{H}(r,p,b)$,
   Compute capacity $c(r,p,b)$ based on SINR$(r,p,b)$
5.             **for** $m = 1 : N_{modulation}$ **do**
6.                 Compute $MI(r,p,b,m)$ based on $c(r,p,b)$ and $m$
7.             Select $\max_m MI(r,p,b,m)$ to yield $MI(r,p,b)$
8.         Compute $MI(r,p) = \frac{1}{N_{RBG}} \sum_{b=1}^{N_{RBG}} MI(r,p,b)$
9.     Select WB RI and TPMI as $r_{OPT}, p_{OPT} = \underset{r,p}{argmax}\, MI(r,p)$
10. **for** $b = 1 : N_{RBG}$ **do**
11.     Select MI per RBG as $MI(r_{OPT}, p_{OPT}, b)$

---

**Algorithm 2: MCS Adaptation Algorithm**

**Input:** $H_{est}$ per RBG, $W$ per RBG, $P$(RI, TPMI), PDO, $\alpha$

**Output:** Adapted MCS per RBG

1. Compute $\alpha H_{est}$ and $\frac{1}{\alpha} H_{est}$
2. **for** $b = 1 : N_{RBG}$ **do**
3.     Compute WB RI, WB TPMI, and MI per RBG using Algorithm 1 based on $H_{est}$, yielding $MI(b, H_{est})$.
4.     Compute MI per RBG using Algorithm 1 based on selected WB RI and WB TPMI, as well as $\alpha H_{est}$ and $\frac{1}{\alpha} H_{est}$, respectively, yielding $MI(b, \alpha H_{est})$ and $MI(b, \frac{1}{\alpha} H_{est})$.
5. Generate a curve of MI vs. reference PDOs as illustrated by the rightmost example plot in Fig. 2.
6. Perform interpolation or extrapolation to obtain an adapted MI corresponding to actual PDO
7. Obtain an adapted MCS via MI2MCS table using the adapted MI

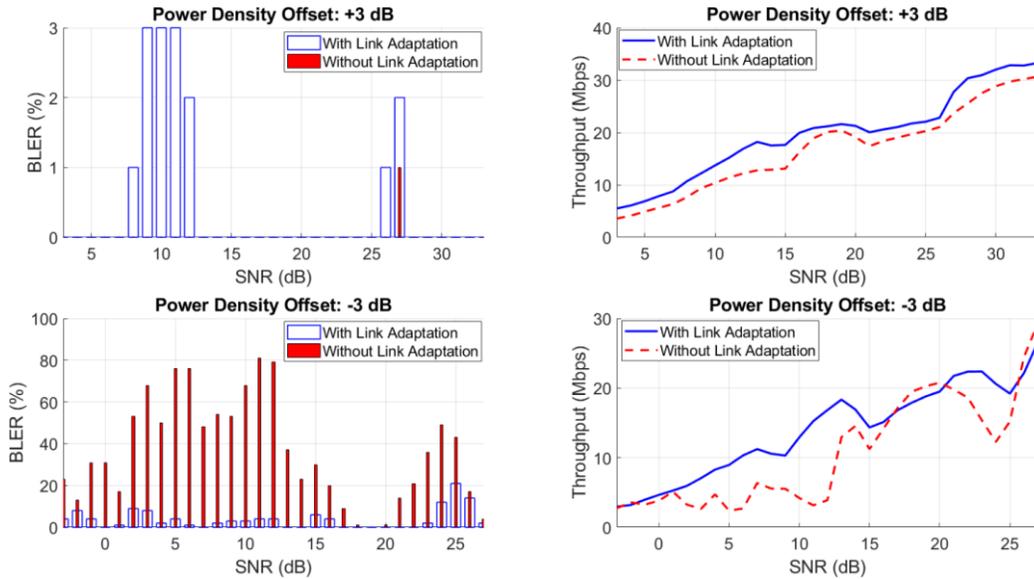

Fig. 3. Comparison of BLERs and throughput between without and with link adaptation cases for PDOs of +3 dB (top) and -3 dB (bottom). The channel is 4x4 EPA5, and the SRS transmission comb is 4.

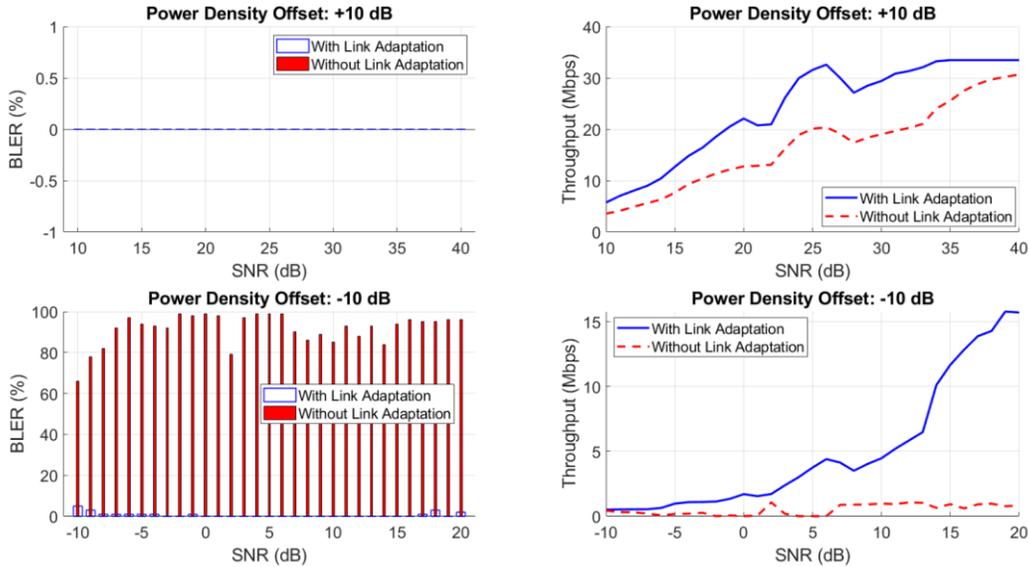

Fig. 4. Comparison of BLERs and throughput between without and with link adaptation cases for PDOs of +10 dB (top) and -10 dB (bottom). The channel is 4x4 EPA5, and the SRS transmission comb is 4.

throughput between without and with link adaptation cases for PDOs of ±3 dB for the 4x4 EPA5 channel are illustrated in Fig. 3, in which the signal-to-noise ratio (SNR) on the x-axis denotes the SNR including the PDO in the PUSCH. It can be observed from Fig. 3 that when no link adaptation is conducted, the BLERs are extremely low (within 1% in this example) for positive 3 dB PDO and extremely high (up to about 80%) for negative 3 dB PDO in the SNR range of 0 dB to 30 dB, due to the lack of MCS adaptation such that the MCS selected for PUSCH transmission is lower than it should be for positive PDOs and higher for negative PDOs. On the other hand, with the proposed link adaptation mechanism in this letter, reasonable BLERs (between 0% and 10%) are maintained at most SNRs between 0 dB and 30 dB for both +3 dB and -3 dB PDOs. Furthermore, as shown by the right two plots of Fig. 3, the proposed link adaptation algorithm yields higher throughput in general for both positive and negative PDOs as compared to the without link adaption case. Fig. 4 depicts the comparison of BLERs and throughput between with and without link adaptation cases for larger PDOs of ±10 dB, which indicates similar performance trends to Fig. 3 from 0 dB to 30 dB SNRs, with even more substantial discrepancies in BLER for -10 dB PDO and throughput for both ±10 dB PDOs between with and without link adaptation cases.

The average throughput gains offered by the proposed link adaptation algorithm compared with no-link-adaptation case are summarized in Table II for ±3 dB and +20 dB PDOs within 0-30 dB SNRs (including the PDO) for two configurations: 4x4 EPA5 channel with transmission comb 4 and 2x2 EVA20 channel with transmission comb 2. Throughput gain is defined herein as the ratio of the difference between throughputs of the two link adaptation cases over the throughput corresponding to the no-link-adaptation case. It is evident from Table II that the throughput gain increases with the PDO. Moreover, similar results are observed for a diversity of types of radio channels, Doppler frequencies, numbers of antenna ports, transmission comb configurations for SRS, etc., which indicates the robustness of our algorithm against various conditions.

Table I Simulation settings

| Parameter | Setting 1 | Setting 2 |
|---|---|---|
| Number of reference PDOs | 3 | 3 |
| Reference PDOs | -10 dB, 0 dB, +10 dB | -10 dB, 0 dB, +10 dB |
| Number of RBs per RBG | 4 | 4 |
| Number of RBGs per UE for SRS | 68 | 68 |
| Number of RBGs per UE for PUSCH | 2 | 2 |
| Transmission comb for SRS | 4 | 2 |
| Radio channel | EPA5 (Extended Pedestrian A model with 5 Hz Doppler frequency) | EVA20 (Extended Vehicular A model with 20 Hz Doppler frequency) |
| Number of transmit antenna ports | 4 | 2 |
| Number of receive antenna ports | 4 | 2 |
| Target BLER | 10% | 10% |

IV. CONCLUSION

In this letter, we have proposed a practical link adaptation algorithm tackling the problem of PDOs between a PUSCH and its associated SRS. Simulation results have demonstrated that our algorithm can effectively keep the BLER within expected values and produce high throughput for a large PDO range of up to ±20 dB, while having low implementation complexity due to constant RI and TPMI over unscaled and scaled channels plus parallel processing as illustrated by Fig. 2 as an example. The throughput gain against no link adaptation can reach over 80% even for a small PDO such as -3 dB and up to 311% for +20 dB PDO. Further variations or improvements, such as more than three reference PDOs, and adaptive instead of fixed reference PDOs, can be applied to make the resultant error rate and/or throughput more satisfactory, with moderate increase in hardware complexity.

Table II Average throughput gain provided by the proposed link adaptation algorithm compared with no-link-adaptation case for 0-30 dB SNRs including the PDO[1].

| PDO (dB) | 4x4 EPA5, comb4 | 2x2 EVA20, comb2 |
|---|---|---|
| +3 | 25% | 32% |
| -3 | 82% | 67% |
| +20 | 177% | 311% |

---

[1] For -20 dB PDO, the throughput gain is nominally infinity because the throughput without link adaptation is 0.